\documentclass[fleqn,usenatbib]{mnras}

\usepackage{newtxtext,newtxmath}
\usepackage{lineno}

\usepackage[T1]{fontenc}

\DeclareRobustCommand{\VAN}[3]{#2}
\let\VANthebibliography\thebibliography
\def\thebibliography{\DeclareRobustCommand{\VAN}[3]{##3}\VANthebibliography}

\usepackage{graphicx}
\usepackage{amsmath}
\usepackage{mathtools}
\usepackage{bm}
\usepackage[dvipsnames]{xcolor}
\usepackage{gensymb}

\newcommand{\rs}{\text{R}_\odot}

\title[Alfv\'en Zone in Young Solar Wind]{The Alfv\'en Transition Zone observed by the Parker Solar Probe in Young Solar Wind -- Global Properties and Model Comparisons}

\author[R. Chhiber et al.]{
Rohit Chhiber$^{1,2}$\thanks{E-mail: rohit.chhiber@nasa.gov},
Francesco Pecora$^2$,
Arcadi V. Usmanov$^{1,2}$,
William H. Matthaeus$^{2,3}$,\newauthor
Melvyn L. Goldstein$^4$, 
Sohom Roy$^2$,
Jiaming Wang$^2$,
Panisara Thepthong$^5$,
and David Ruffolo$^6$
\\ \\
$^{1}$Heliophysics Science Division, NASA Goddard Space Flight Center, Greenbelt, MD 20771, USA\\
$^{2}$Department of Physics and Astronomy, University of Delaware, Newark, DE 19716, USA\\
$^{3}$Bartol Research Institute, University of Delaware, Newark, DE 19716, USA\\
$^4$Space Science Institute, Boulder, CO 80301, USA\\
$^5$Department of Physics, Faculty of Science, Kasetsart University, Bangkok 10900, Thailand\\
$^6$Department of Physics, Faculty of Science, Mahidol University, Bangkok 10400, Thailand 
}

\date{Accepted XXX. Received YYY; in original form ZZZ}

\pubyear{2024}

\begin{document}
\label{firstpage}
\pagerange{\pageref{firstpage}--\pageref{lastpage}}
\maketitle

\begin{abstract}
The transition from subAlfv\'enic to superAlfv\'enic flow in the solar atmosphere is examined by means of Parker Solar Probe (PSP) measurements during solar encounters 8 to 14. Around 220 subAlfv\'enic periods with a duration \(\ge\) 10 minutes are identified. The distribution of their durations, heliocentric distances, and Alfv\'en Mach number are analyzed and compared with a global magnetohydrodynamic model of the solar corona and wind, which includes turbulence effects. The results are consistent with a patchy and fragmented morphology, and suggestive of a  turbulent Alfv\'en zone within which the transition from subAlfv\'enic to superAlfv\'enic flow occurs over an extended range of helioradii. These results inform and establish context for detailed analyses of subAlfv\'enic coronal plasma that are expected to emerge from PSP's final mission phase, as well as for NASA's planned PUNCH mission. 
\end{abstract}




\section{Introduction}


In its eighth orbit, the Parker Solar Probe (PSP) first encountered periods of sub-Alfv\'enic flow at a heliocentric distance of about 16 $R_\odot$
\citep{kasper2021prl}.
The manifold implications of this transition -- from plasma flow speeds less than to greater than the Alfv\'en speed -- have been widely discussed \citep{Cranmer2023SoPh}, and are intertwined with the PSP mission's primary science goals \citep{fox2016SSR}. Not only is a
transition at a well defined surface
related to a simple expression
for the rate of angular momentum loss of the sun \citep{weber1967ApJ148}, but 
it also separates regions in which the characteristics of 
magnetohydrodynamic waves are expected to behave quite differently \citep{verdini2009ApJ,Cranmer2023SoPh}. 
Notionally such a surface also represents 
a separation, though not a true boundary, between, what is commonly thought of as ``solar wind'' above, and ``corona'' below.  
In this regard it is noteworthy that 
evidence is accumulating \citep{bandyopadhyay2022ApJ,Zhao2022ApJL_subA1,Zhao2022ApJL_subA2,Zank2022ApJ,Jiao2024ApJ} that suggests measurable differences in plasma and turbulence properties
when organized by this criterion. Recent work suggests that the Alfv\'enic transition may also play a key role in the onset of shear-driven instability and enhanced turbulence \citep{deforest2016ApJ828,ruffolo2020ApJ}.

The location of this transition is usually called the Alfv\'en \textit{point} or \textit{radius} in a one-dimensional context, and the Alfv\'en \textit{surface} in three-dimensional (3D) representations obtained from global magnetohydrodynamic (MHD) solar wind models \citep[e.g.,][]{chhiber2019psp1}. While large-scale models depict this surface as a smooth manifold, recent studies have suggested that the surface may be highly irregular and ``corrugated'' \citep{liu2021ApJ,wexler2021ApJ,Verscharen2021MNRAS,Jiao2024ApJ}, or even 
a ``zone'' consisting of interspersed
patches of super- and sub-Alfvenic plasma regions \citep[][see below]{deforest2018ApJ,Chhiber2022MNRAS}. With the ever increasing number of PSP entries into subAlfv\'enic plasma (which we view, presumptively, as coronal plasma) since the first observations in 2021, it is now feasible to further investigate the nature of this transition using in situ measurements. The present paper initiates a step in this direction by examining certain global properties of the observed subAlfv\'enic periods, and along with comparisons with a model Alfv\'en zone. Our principal goal is to understand whether the observations continue to support the idea that the Alfv\'enic transition is complex, and neither smooth nor monotonic. The conclusion  will be in the affirmative, although we will not claim that our findings are conclusive.
The present results will also serve to establish context for more detailed analyses that are anticipated from PSP's final mission phase.

\section{PSP Data and Identification of SubAlfv\'enic Intervals}\label{sec:data}

PSP measurements from solar encounters (denoted `E') 8 to 14 are used to study  properties of subAlfv\'enic intervals, covering heliocentric distances from \(\sim\) 13-40\(~\rs\) during the time period from March 2021 to Dec 2023. Recall that PSP first measured subAlfv\'enic solar wind during E8 \citep{kasper2021prl}. Magnetic field measurements by the fluxgate magnetometer (MAG) on the FIELDS suite \citep{bale2016SSR} are used, while plasma measurements are from the SPAN-I instrument aboard the SWEAP suite \citep{kasper2016SSR,Livi2022ApJ}. SPAN provides better coverage of the plasma distribution near the Sun compared to the SPC instrument, in the range of helioradii of primary interest for this study. Vector quantities are expressed in a heliocentric \(RTN\) coordinate system \citep{franz2002pss}.

The data are smoothed to a 1-minute cadence; this (relatively coarse) choice is motivated by our focus on the macroscopic morphology of subAlfv\'enic parcels of solar wind, rather than detailed variations within such parcels. To ensure that we identify coherent subAlfv\'enic parcels unaffected by small-scale fluctuations in the Alfv\'en Mach number, our reference scale is based on estimates of correlation times of magnetic fluctuations during PSP encounters; these tend to be of the order of several minutes, with shorter correlation times observed for decreasing heliocentric distance \(r\) \citep[e.g.,][]{chen2020ApJS,parashar2020ApJS,Cuesta2022ApJL}. The 1-min cadence data then represent the average plasma and magnetic field properties at approximately the correlation scale.\footnote{For reference, the ion gyrofrequency at PSP perihelia lies at scales smaller than 1~s \citep[e.g.,][]{Chhiber2021ApJL}.} For most of our analysis (see below), we further require identified subAlfv\'enic intervals to have a sustained duration of at least 10 minutes.

Our identification of subAlfv\'enic intervals is based on the following definition of the Alfv\'en Mach number:
\begin{equation}
    M_A = V_R/V_A,
\end{equation}
where \(V_R\) is the radial component of the bulk ion velocity and \(V_A = B/\sqrt{4\pi\rho}\) is the Alfv\'en speed, computed from the magnetic field magnitude \(B\) and the ion density \(\rho\). The 1-min cadence \(M_A\) time series is then scanned for periods when \(M_A < 1\) \textit{continuously}. With the further requirement that the identified intervals be at least 10 minutes long, we obtain 228 subAlfv\'enic intervals from E8-E14.

\section{Model Alfv\'en Zone}\label{sec:model}

\begin{figure}
    \centering
    \includegraphics[width=.8\columnwidth]{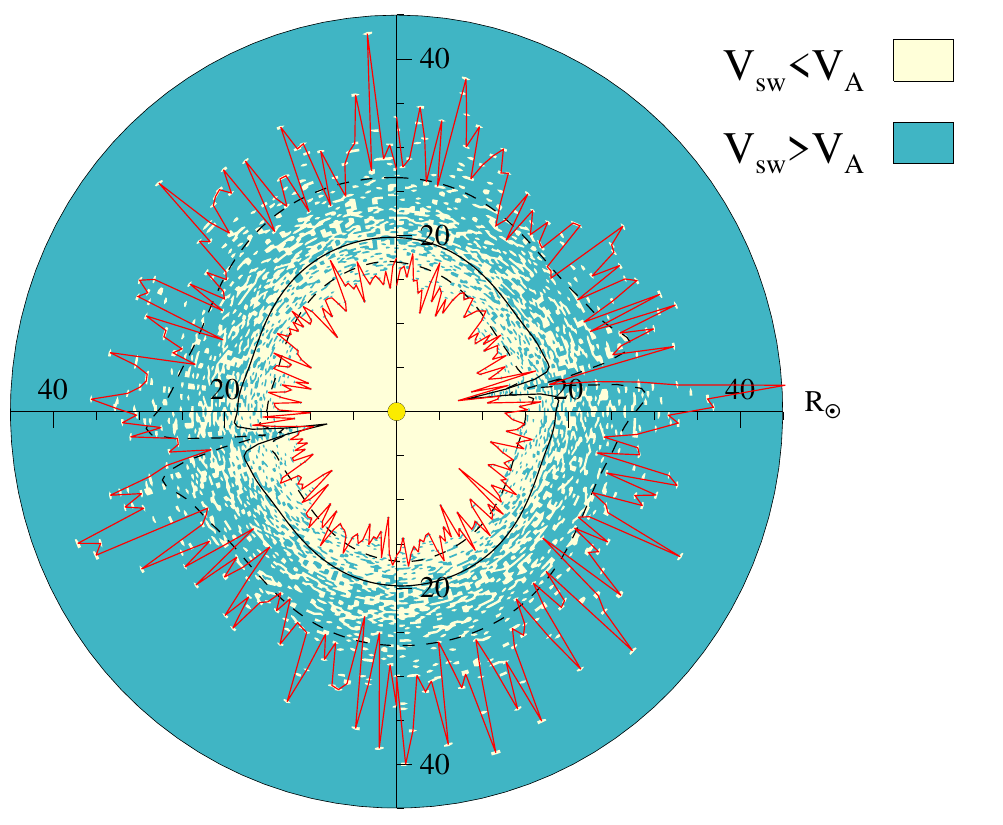}
    \caption{Model Alfv\'en zone. SubAlfv\'enic (beige) and superAlfv\'enic (teal) regions are distinguished by colour in meridional plane from \(10\degree\)-tilt dipole simulation. Helioradii from 1 to \(45~\rs\)~are shown. Solid black curve shows Alfv\'en surface computed from mean fields. Inner red curve marks first super-Alfv\'enic (teal) point while moving outward along a radial spoke. Outer red curve marks last subAlfv\'enic (beige) point while moving outward along a radial spoke. See \protect\cite{Chhiber2022MNRAS} for more details.}    
    \label{fig:azone}
\end{figure}

A possible realization of an extended and fragmented Alfv\'en zone was presented by \cite{Chhiber2022MNRAS}, using a 3D MHD model of the global solar wind
which includes self-consistent turbulence transport \citep{usmanov2018}. This model provides mean-field (or bulk flow) solar wind parameters throughout the inner heliosphere, including proton velocity and density and magnetic field, as well as statistical descriptors of turbulence, including average fluctuation energy, cross helicity, and a correlation scale. 
Our approach was based on generating a realization of explicit magnetic fluctuations having a random character but constrained by the 3D distribution of rms turbulence amplitudes provided by the model. The addition of this fluctuation field to the mean magnetic field produced the extended and fragmented Alfv\'en zone. A sample representation in a meridional plane is shown in Figure \ref{fig:azone}; this result is based on a simulation with a source magnetic dipole tilted by \(10\degree\) relative to the solar rotation axis. For more details see \cite{Chhiber2022MNRAS}.

Initial comparisons of the Alfv\'en zone model with PSP observations have shown good agreement \citep{Chhiber2022MNRAS,Cranmer2023SoPh}. 
While observations by a constellation of spacecraft would be ideal for distinguishing between the Alfv\'en ``surface'' vs ``zone'' pictures, even for a single spacecraft \cite{Chhiber2022MNRAS} predicted that the zone picture ``implies greater frequency-of-occurrence of subAlfv\'enic patches, and longer durations as well, as PSP descends to lower perihelia. 
The surface picture implies longer durations of subAlfv\'enic intervals as PSP descends, instead of increasing ... frequency of such intervals''. 
Below we will further compare the properties of subAlfv\'enic intervals observed by PSP with the Alfv\'en zone model. Our goal is to examine global distributions of subAlfv\'enic parcels in this region, aggregating data from several PSP encounters; so, rather than detailed comparisons of individual encounters with model runs based on specific Carrington Rotation magnetograms \citep[as in][]{chhiber2021ApJ_psp}, we will use a single representative model run based on a solar magnetic dipole \citep[Run I of][]{Chhiber2022MNRAS}.

\section{Results}\label{sec:results}

\begin{figure}
    \centering
    \includegraphics[width=.9\columnwidth]{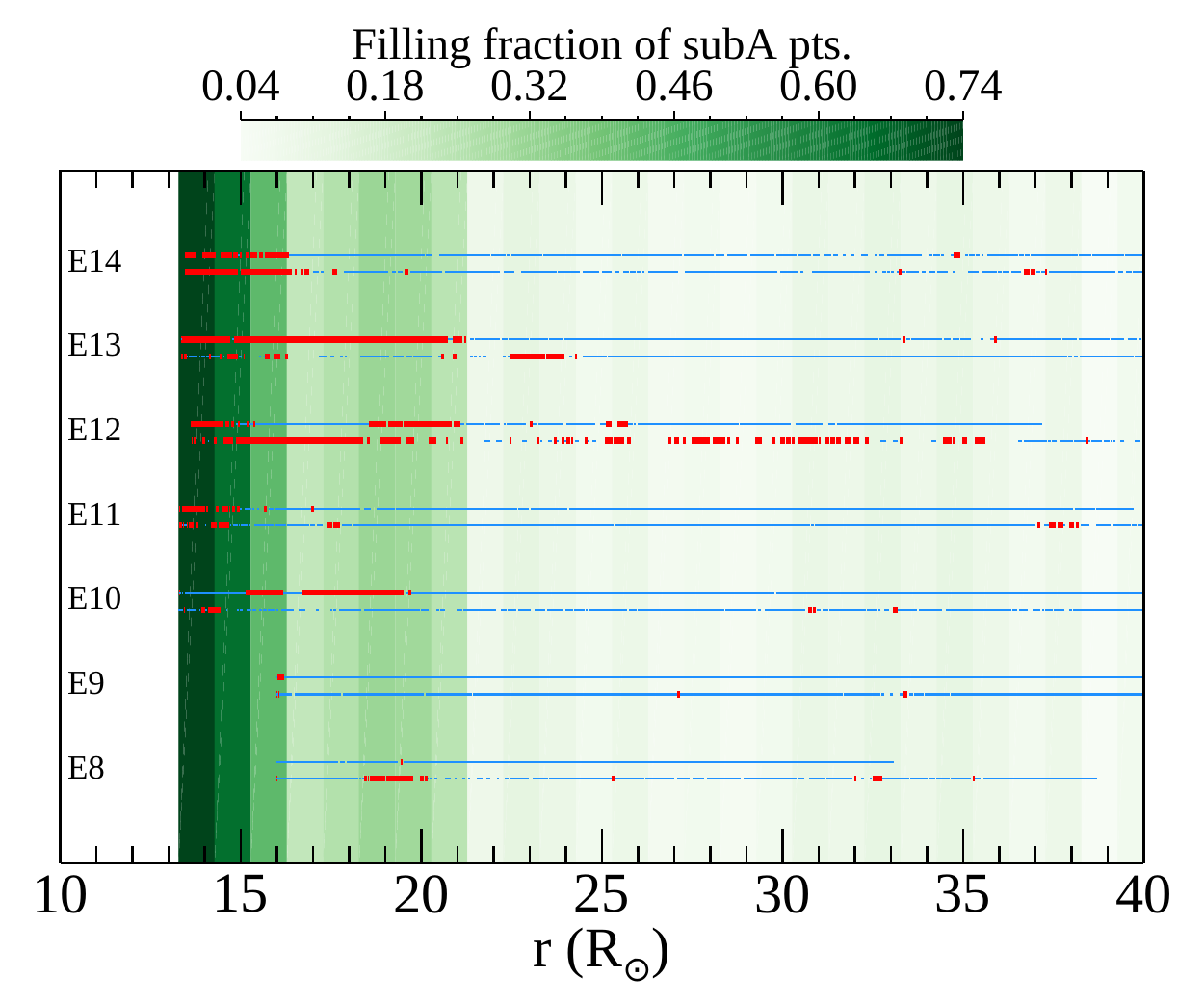}
    \includegraphics[width=.85\columnwidth]{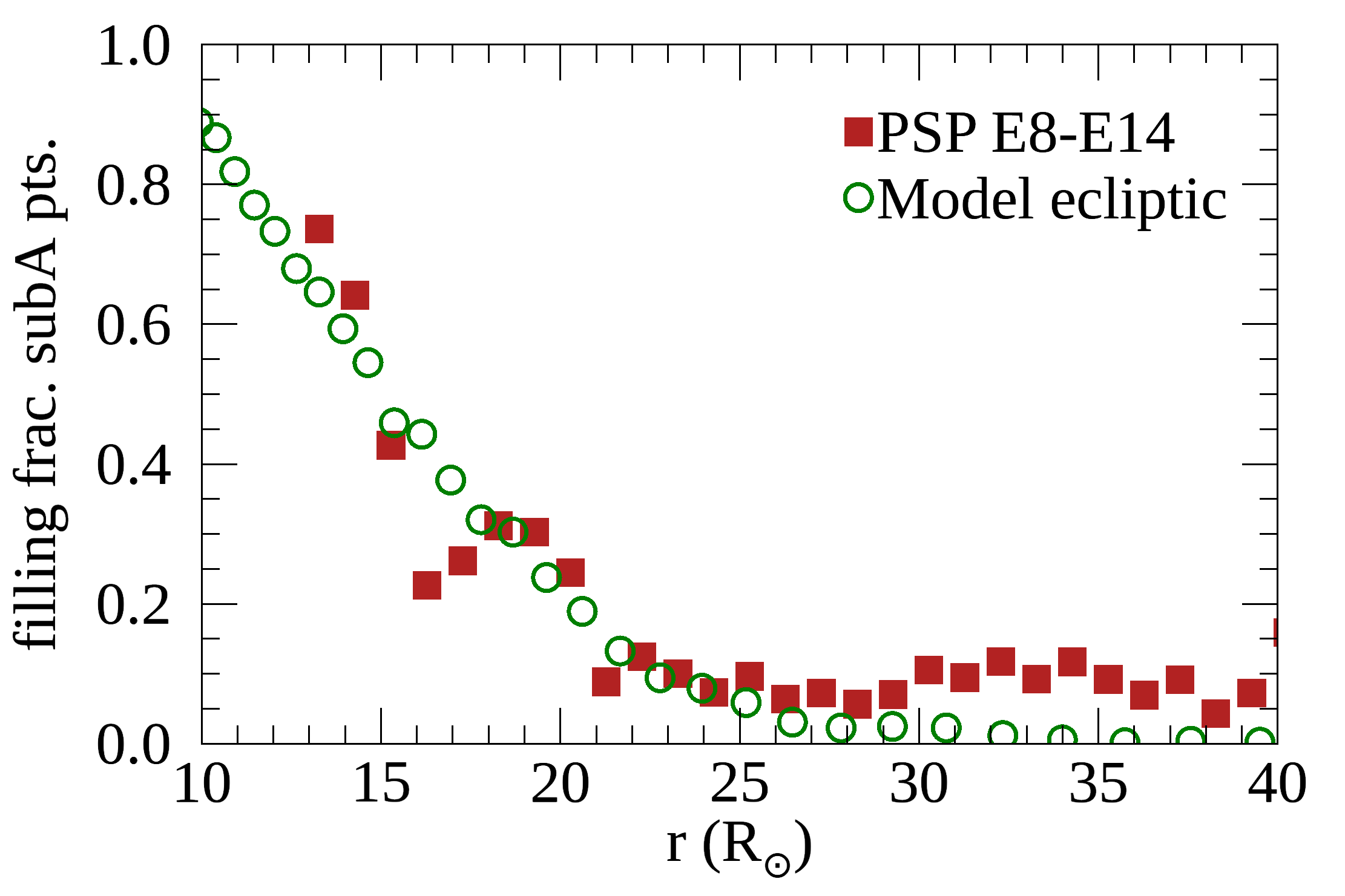}
    \caption{\textit{Top}: Distribution of subAlfv\'enic and superAlfv\'enic intervals measured by PSP during E8 to E14, as function of helioradius. Each encounter is represented as two proximate horizontal lines, starting with E8 at the bottom and proceeding upward as labeled. For each encounter, bottom horizontal line shows inbound (Sunward) section of PSP's trajectory, while  upper horizontal line shows outbound section. Red and blue segments of horizontal lines denote subA and superA intervals, respectively, with minimum duration \(\ge\) 10 minutes. Gaps in horizontal lines come from periods when no subA or superA intervals longer than 10 minutes were present. Vertical shaded columns denote radial bins of width \(5~\rs\), with color depicting filling fraction (see text) of subAlfv\'enic points within a bin, aggregated across all shown encounters. \textit{Bottom}: Filling fraction of subAlfv\'enic points from model (green circles) and PSP (brown squares), as function of helioradius. PSP data from E8-E14 are aggregated in radial bins of width \(1~\rs\). Model calculation includes data from all heliolongitudes and \(\pm 6\degree\) in heliolatitude, which are aggregated in \(1~\rs\) radial bins.
    } \label{fig:morse_ff}
\end{figure}

\begin{figure}
    \centering
    \includegraphics[width=.85\columnwidth]{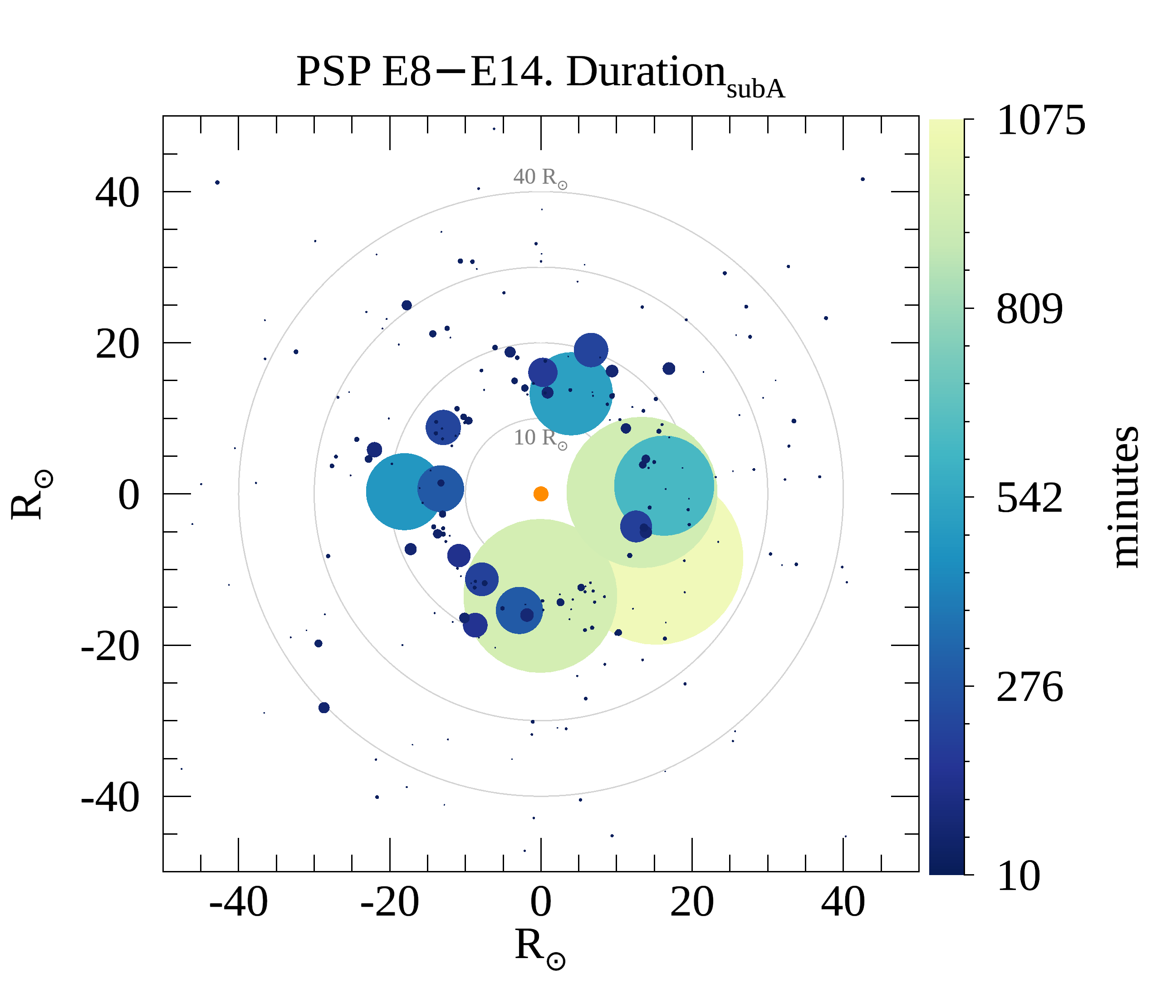}
    \includegraphics[width=.85\columnwidth]{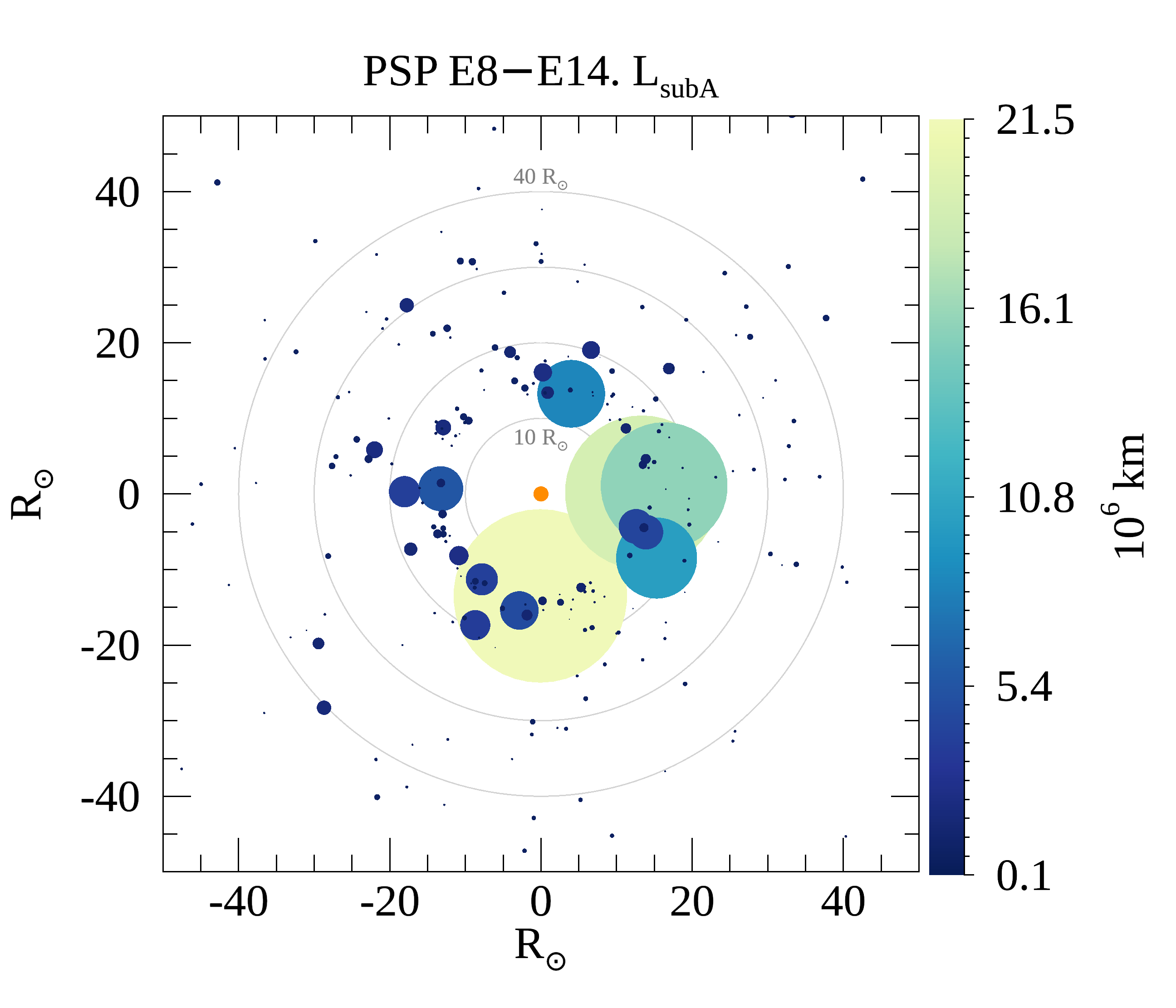}
    \includegraphics[width=.85\columnwidth]{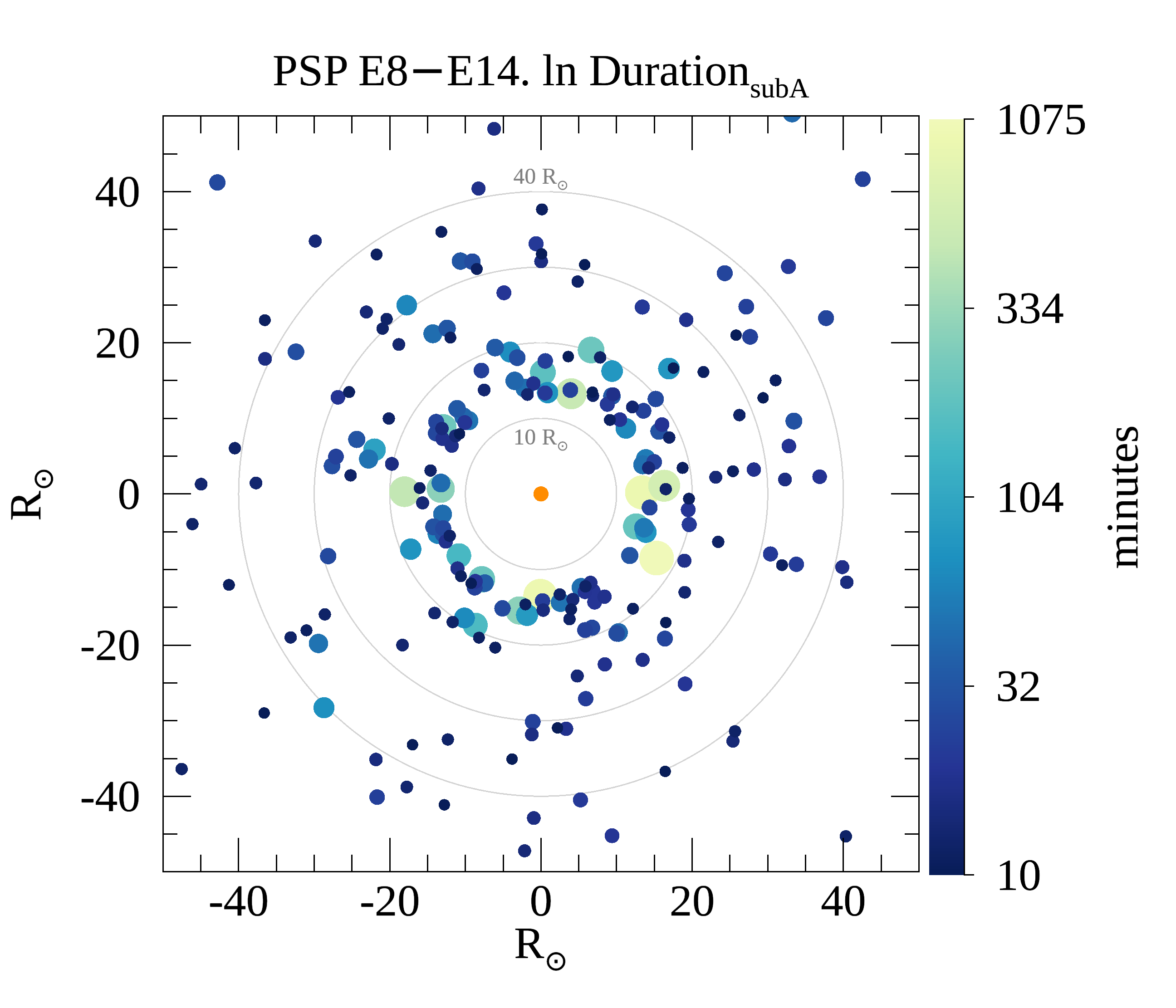}
    \caption{Bubble plots showing time and length scales of subAlfv\'enic intervals in PSP E8-E14, in ecliptic region corresponding to PSP's trajectory. Each circle represents a subA interval with a duration \(\ge\) 10 min, where center of circle corresponds to helioradius of corresponding interval. Longitude of each interval is assigned a random value from a uniform distribution between 0\degree~and 359\degree~(see text). Diameters of circles represent quantity labeled at top of each panel, and are scaled proportionally to the smallest circle, which has arbitrary diameter. Colorbar also represents respective labeled quantities, which are, from top to bottom, duration, size \(L_\text{subA}\) as defined in text, and natural log of duration. Central orange circle represents Sun (to scale with axes) and concentric grey circles show reference heliodistances of 10, 20, 30, and 40 \(\rs\).}
    \label{fig:bub1}
\end{figure}

In the following we sometimes use ``subA'' and ``superA'' as acronyms for ``subAlfv\'enic'' and ``superAlfv\'enic'', respectively. 
We first examine observed distributions of subAlfv\'enic intervals along the PSP trajectory, as a function of \(r\). The top panel of Figure \ref{fig:morse_ff} depicts subAlfv\'enic and superAlfv\'enic intervals with a minimum duration of 10 minutes as thick red and thin blue segments, respectively, along horizontal lines that represent inbound (Sunward) and outbound sections of each solar encounter separately. The occurrence of subAlfv\'enic segments increases with decreasing \(r\), as does their apparent spatial extent. The patchy or corrugated nature of the transition from subA to superA wind is also apparent, with the two types of segments interspersed with each other. The \textit{apparent} spatial extent (in the \(r\) direction) of continuous subA parcels ranges from a fraction of a solar radius to more than \(5~\rs\) (or \(\sim 4\times 10^6\) km). Note that this apparent size depends on the relative motion of PSP and the solar wind parcel, and the actual spatial extent of the parcels is expected to be larger (see Figure \ref{fig:bub1}). Complementary depictions of the PSP trajectory that show similarly patchy \textit{longitudinal} distributions of subA intervals have been shown in \cite{badman2023JGR}.

Shaded vertical columns in the top panel of Figure \ref{fig:morse_ff} show the filling fraction of subA points in the 1-min cadence PSP time series: data from E8 to E14 are aggregated in \(1\rs\)-wide radial bins, and the filling fraction in a bin is the number of points with \(M_A < 1\) divided by the total number of points. Note that this calculation is not restricted to the 10+ min long intervals in order to obtain larger statistical samples per bin. The filling fraction increases from below 0.1 at \(40~\rs\) to  0.7 at \(14~\rs\). In the bottom panel of Figure \ref{fig:morse_ff} we compare the observed filling fraction with one computed from the model. The model calculation includes data from all heliolongitudes, but only within the ecliptic region (\(\pm 6\degree\) heliolatitude) traversed by PSP. The filling fraction is then computed by aggregating these data within \(1~\rs\) wide radial bins. PSP data at 1-min cadence from E8-E14 are also grouped in \(1~\rs\) bins in this case. Good agreement is seen between the model and observations, especially considering that the model run is based on a generic solar dipole magnetic field rather than ``bespoke'' magnetogram-based runs corresponding to individual PSP encounters. PSP data have more variability than the model, which is reflected in the finite filling fraction seen in PSP above \(30~\rs\).

\begin{figure}
    \centering
    \includegraphics[width=.9\columnwidth]{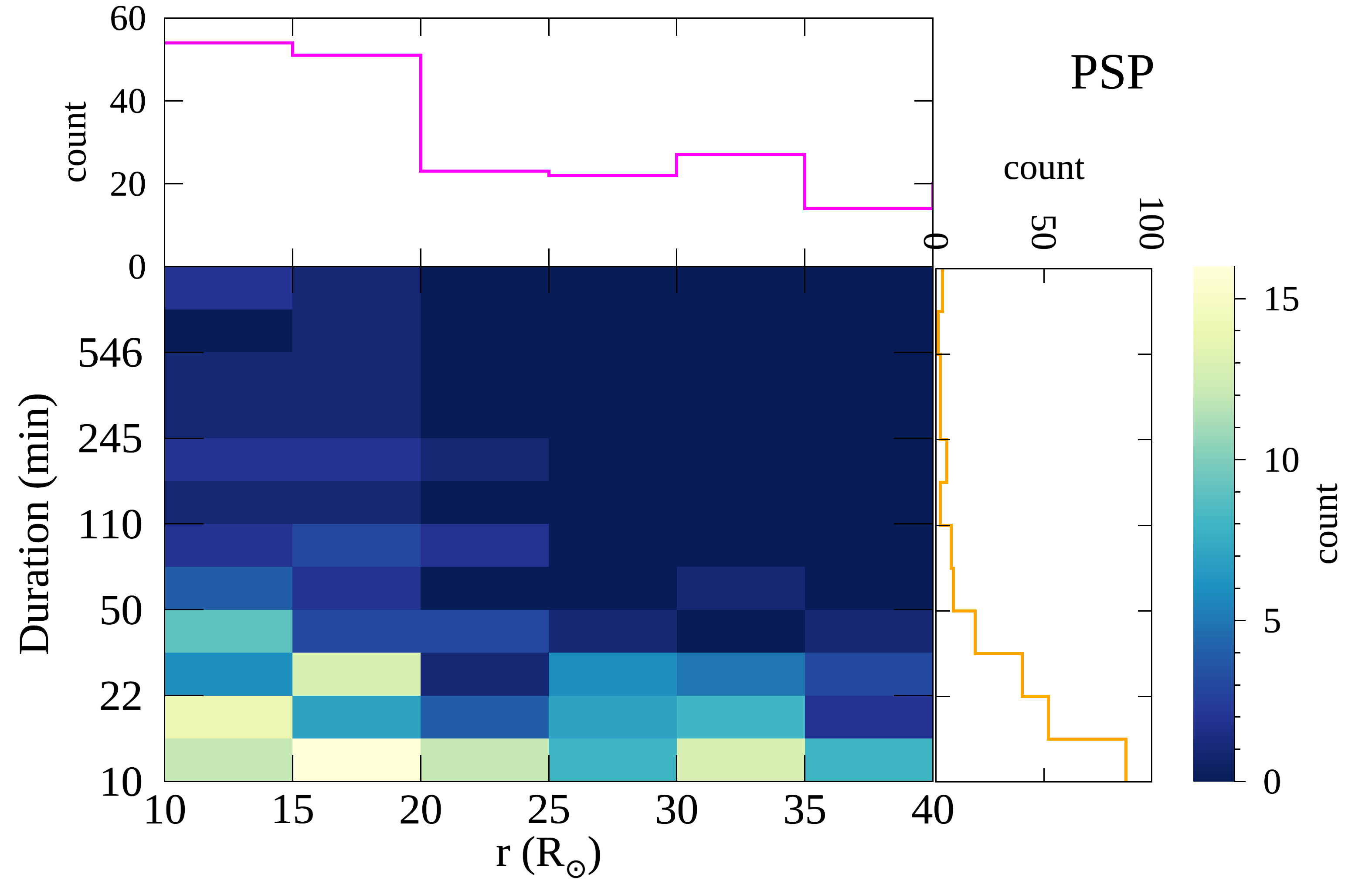}
    \includegraphics[width=.9\columnwidth]{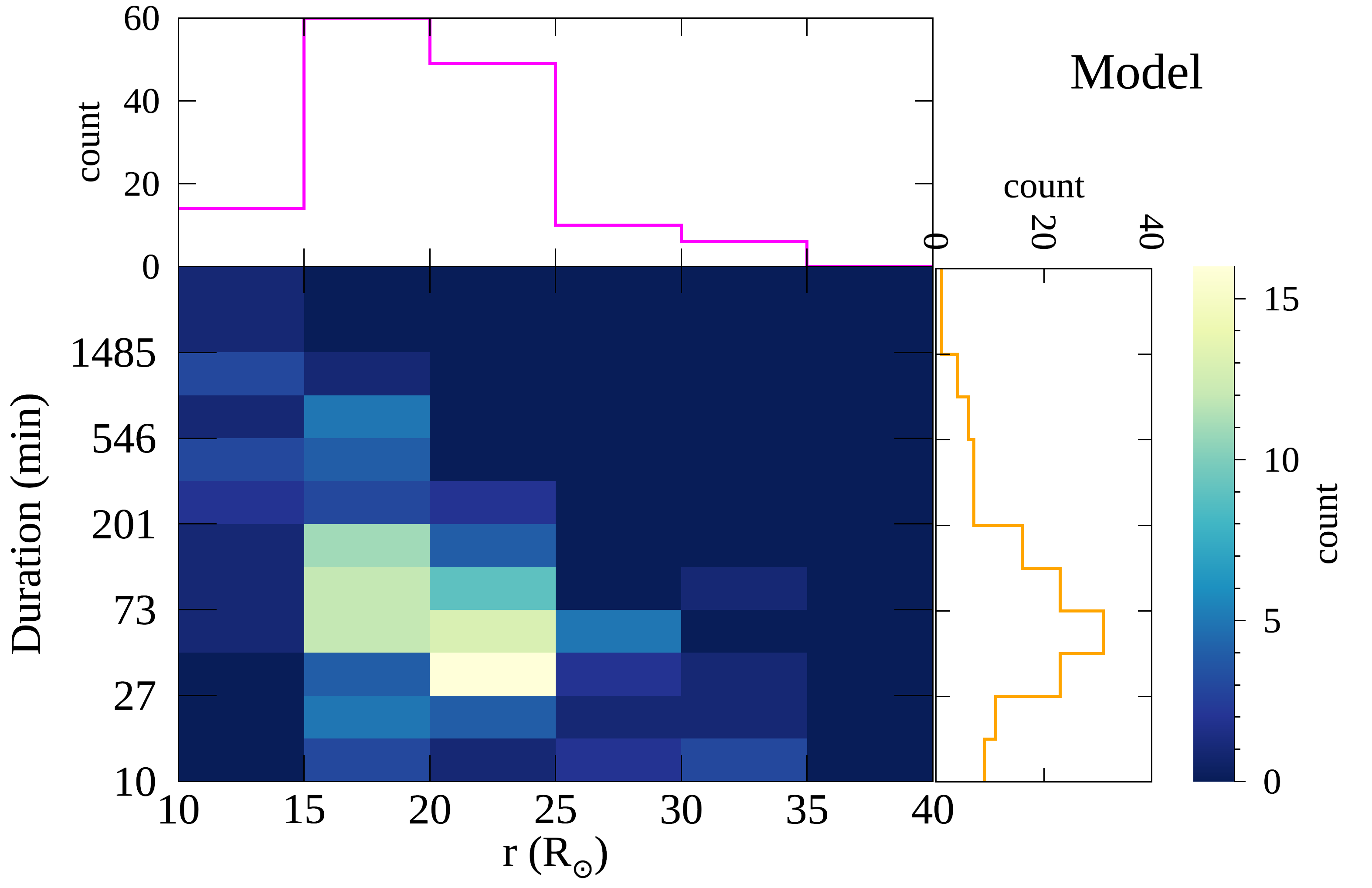}
    \includegraphics[width=.9\columnwidth]{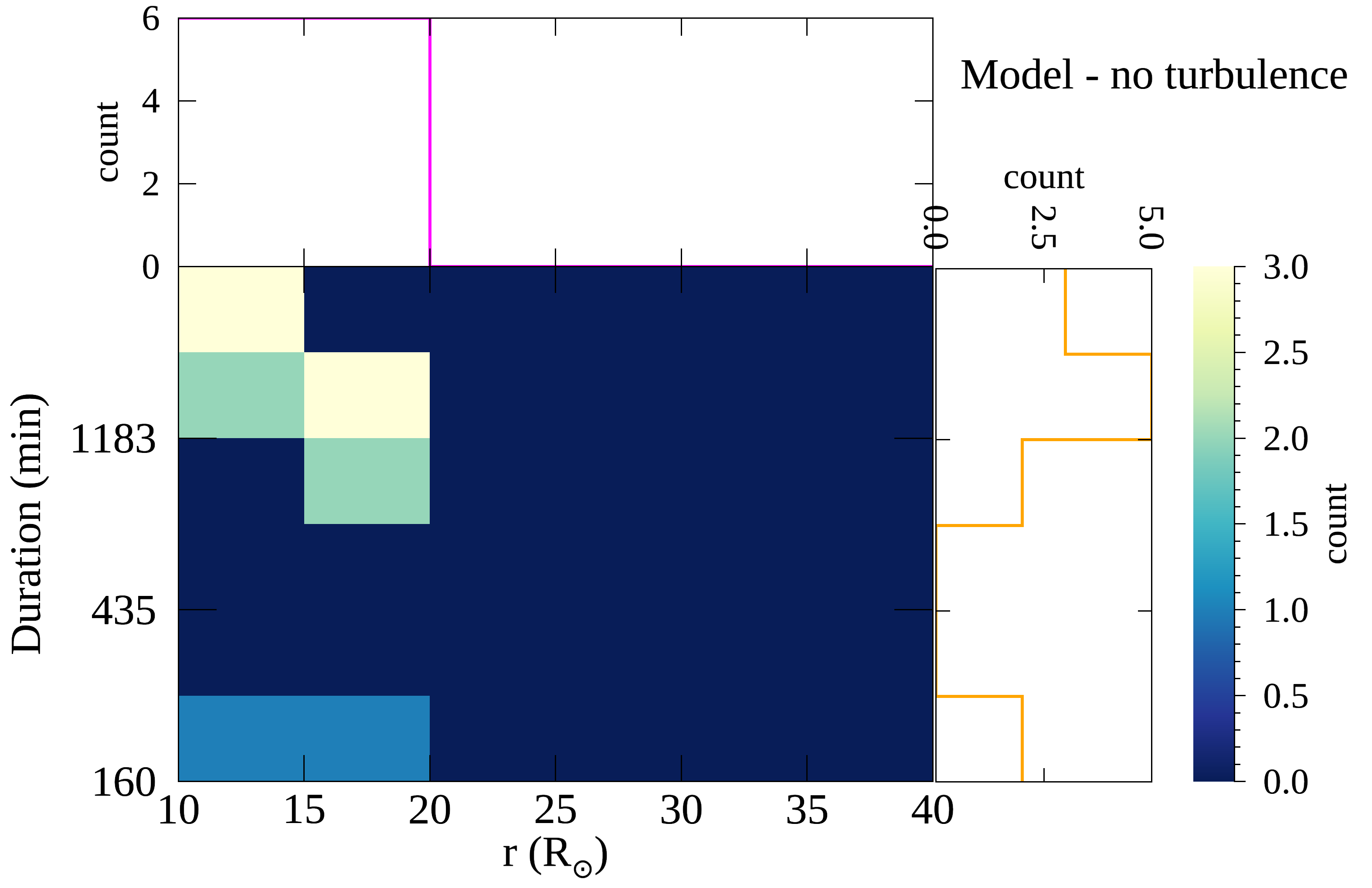}
    \caption{\textit{Top}: 2D histogram of subAlfv\'enic intervals in PSP E8-E14, showing interval count as function of helioradius \(r\) and interval duration, for intervals with minimum duration of 10 minutes. Top sub-panel (magenta curve) shows interval count as a 1D function of \(r\), integrating along duration axis. Right sub-panel (orange curve) shows interval count as function of duration, integrating along \(r\) axis. \textit{Middle}: Analogous plot from model, generated by interpolating model results to ``virtual'' PSP trajectory (see text).  \textit{Bottom}: Analogous plot from model without turbulence (see text). In all panels radial bins are \(5~\rs\) wide and bins along duration axis are logarithmically spaced.} 
    \label{fig:2dhist_1}
\end{figure}

We move now to quantify the duration and occurrence rate of subAlfv\'enic intervals at different \(r\). The top panel of Figure \ref{fig:2dhist_1} presents joint histograms of duration and helioradius of 10+ minute long subA intervals observed by PSP during E8-E14. Attached to the 2D histogram are two sub-panels showing 1d histograms of \(r\) and duration (top and right sub-panels, respectively). Before binning in \(r\), each interval is assigned a radial position which corresponds to the temporal midpoint of the interval. A number of trends are revealed -- (i) The likelihood of long intervals increases approaching the Sun. Durations longer than an hour are typically seen only below \(20~\rs\). (ii) Overall interval count increases approaching the Sun. (iii) Frequency of short duration (10-20 min) intervals appears to increase with decreasing \(r\); that is, even as PSP \textit{begins} to see long subA intervals approaching the Sun, it \textit{continues} to see short duration intervals, instead of the former replacing the latter. This observation is consistent with the prediction of \cite{Chhiber2022MNRAS} in favor of an Alfv\'en zone picture as opposed to an Alfv\'en surface.

For a rough comparison of PSP observations with our model, we fly a virtual PSP trajectory (with a temporal cadence of 10 minutes) for orbits 8 to 14 through the 3D model domain, linearly interpolating the solar wind radial velocity and the Alfv\'en speed to the trajectory, and thus computing the model \(M_A\) along the trajectory. The procedure described in Section \ref{sec:data} is then used to identify subA intervals from the model, and the resulting histograms of duration and \(r\) are shown in the middle panel of Figure \ref{fig:2dhist_1}. The overall trend is consistent with the top panel, with some notable differences: (i) The model has very few subA intervals above \(30~\rs\), in contrast to PSP, which is not surprising given the higher variability contained in PSP measurements (or equivalently, the actual solar wind). (ii) The model has very few intervals in the smallest duration bins (10 - 27 min).  This is probably because the model time series has a 10-minute cadence whereas in the PSP case, the time series cadence is 1 min; we choose a 10 minute cadence for the model PSP trajectory since that roughly corresponds to the model's spatial grid resolution (a few correlation scales). (iii) The lowest radial bin  (10-15\(~\rs\)) in the model panel has almost no intervals shorter than an hour. This is likely because the general radial position of the Alfv\'en zone is at a slightly larger height in the model compared to PSP. 
Indeed, below \(15~\rs\) the virtual PSP trajectory ventures deep within the Alfv\'en zone, close to the region of purely subA wind (see Figure \ref{fig:azone}), which manifests as long durations along the trajectory. Alternatively, the high occurrence rate of short-duration subA intervals at low \(r\) in PSP may indicate that the observed Alfv\'en zone is \textit{more} fragmented and ``patchy'' than in the model.

We note that the values of the durations of subA intervals are roughly similar between PSP and the model, in spite of the model being based on a generic solar dipole rather than solar magnetograms corresponding to the specific PSP encounters under study. One may expect the level of agreement to increase in a more elaborate comparison using a separate simulation run corresponding to each of E8 to E14, using the respective appropriate magnetograms, and then aggregating data from these runs.

The bottom panel of Figure \ref{fig:2dhist_1} shows histograms of subA intervals along the virtual PSP trajectory for a model in which the conventional Alfv\'en surface is computed from the mean fields, without the addition of turbulence \citep[solid black curve in Figure \ref{fig:azone}; see][]{Chhiber2022MNRAS}. The difference between this result and the PSP result is stark, and reinforces the importance of turbulence in developing realistic models of the heliosphere \citep{matthaeus2011SSR,miesch2015SSR194}.

\begin{figure}
    \centering
    \includegraphics[width=.9\columnwidth]{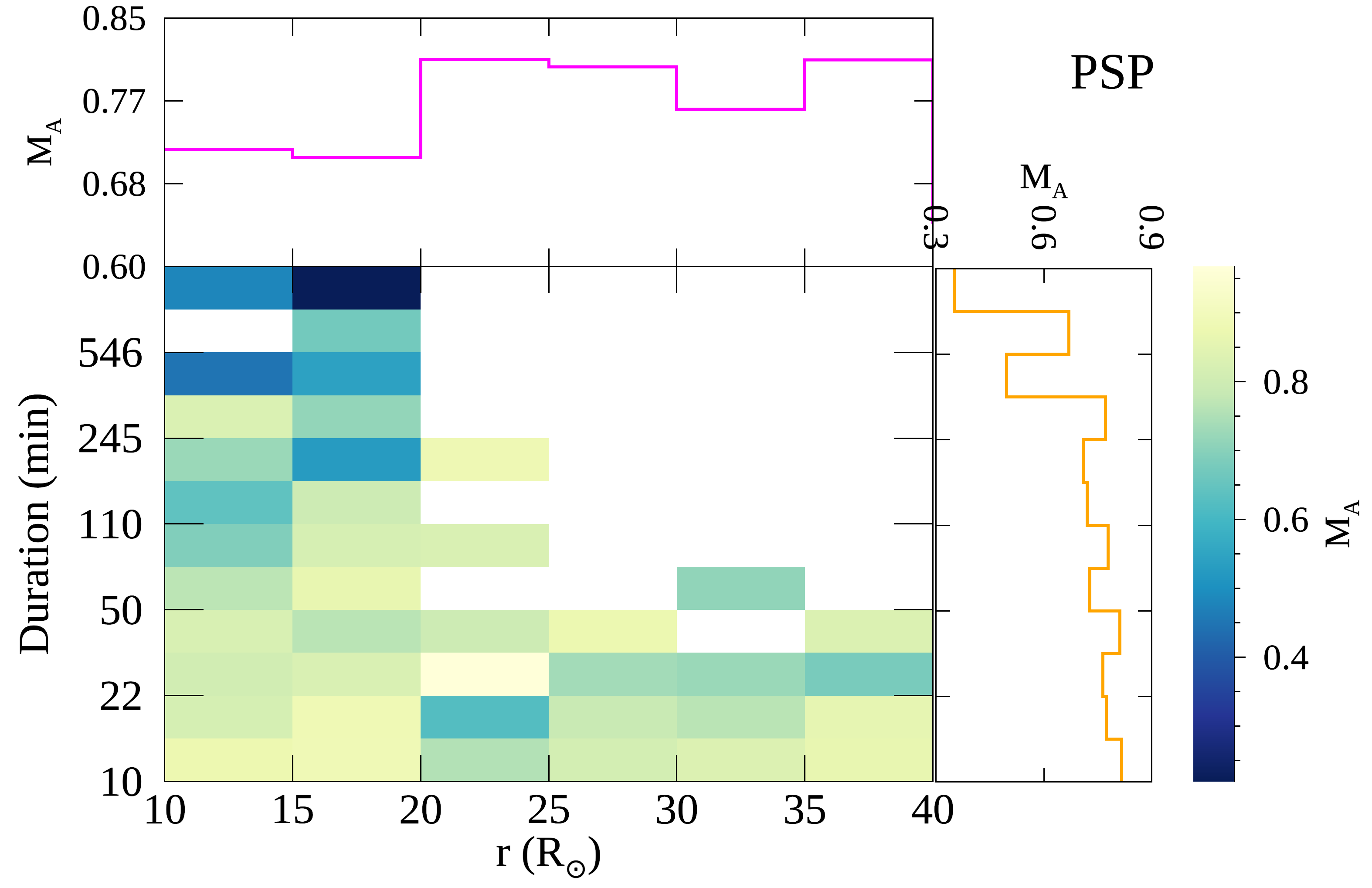}
    \includegraphics[width=.9\columnwidth]{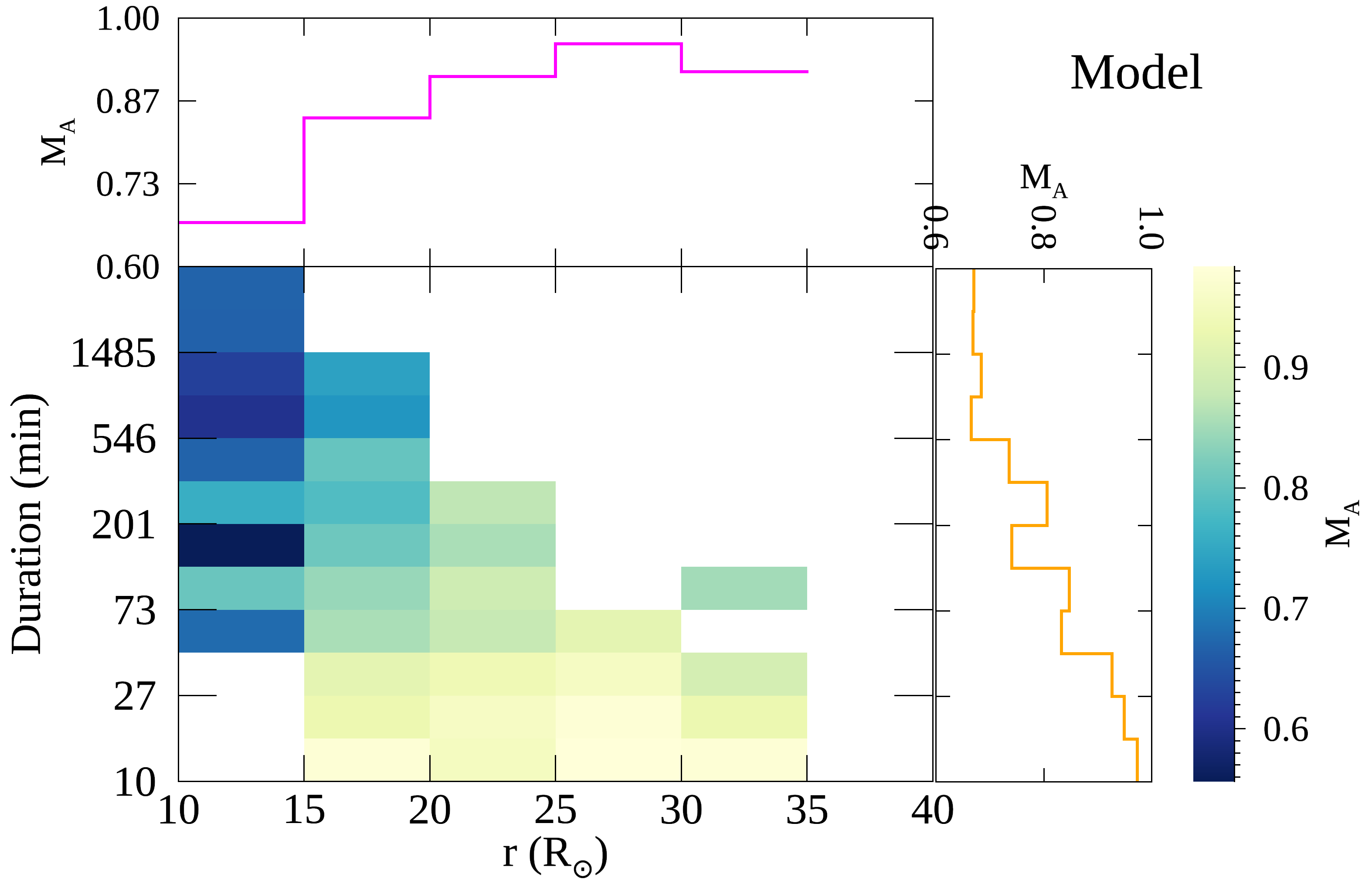}
    \caption{Top: Joint distribution of Alfv\'en Mach number \(M_A\) of subAlfv\'enic intervals as a function of \(r\) and interval duration, in PSP E8-E14, for intervals with minimum duration of 10 minutes. Sub-panels show 1D distributions of \(M_A\), separately as functions of \(r\) and duration. Middle: Analogous plot from model, generated by interpolating model results to a ``virtual'' PSP trajectory (see text). In all panels radial bins are \(5~\rs\) wide and bins along vertical axis are logarithmically spaced.} 
    \label{fig:2dhist_2}
\end{figure}

Next, we examine distributions of the Alfv\'en Mach number in subAlfv\'enic intervals observed in PSP and the model, presented in Figure \ref{fig:2dhist_2} in a format similar to Figure \ref{fig:2dhist_1}. The displayed values of \(M_A\) represent average values across all intervals that lie within the respective bins of \(r\) and/or duration. The same overall trends are observed in both cases. \(M_A\) of subA intervals decreases approaching the Sun, with the model showing a more pronounced decrease. Longer intervals tend to have lower \(M_A\), which is related to their occurrence at lower \(r\). We note again the absence of short-duration subA intervals in the model below 15 \(\rs\). We also note the singular very-long duration PSP interval with an unusually low \(M_A\sim 0.25\).

\begin{figure*}
    \centering
    \includegraphics[width=.51\columnwidth]{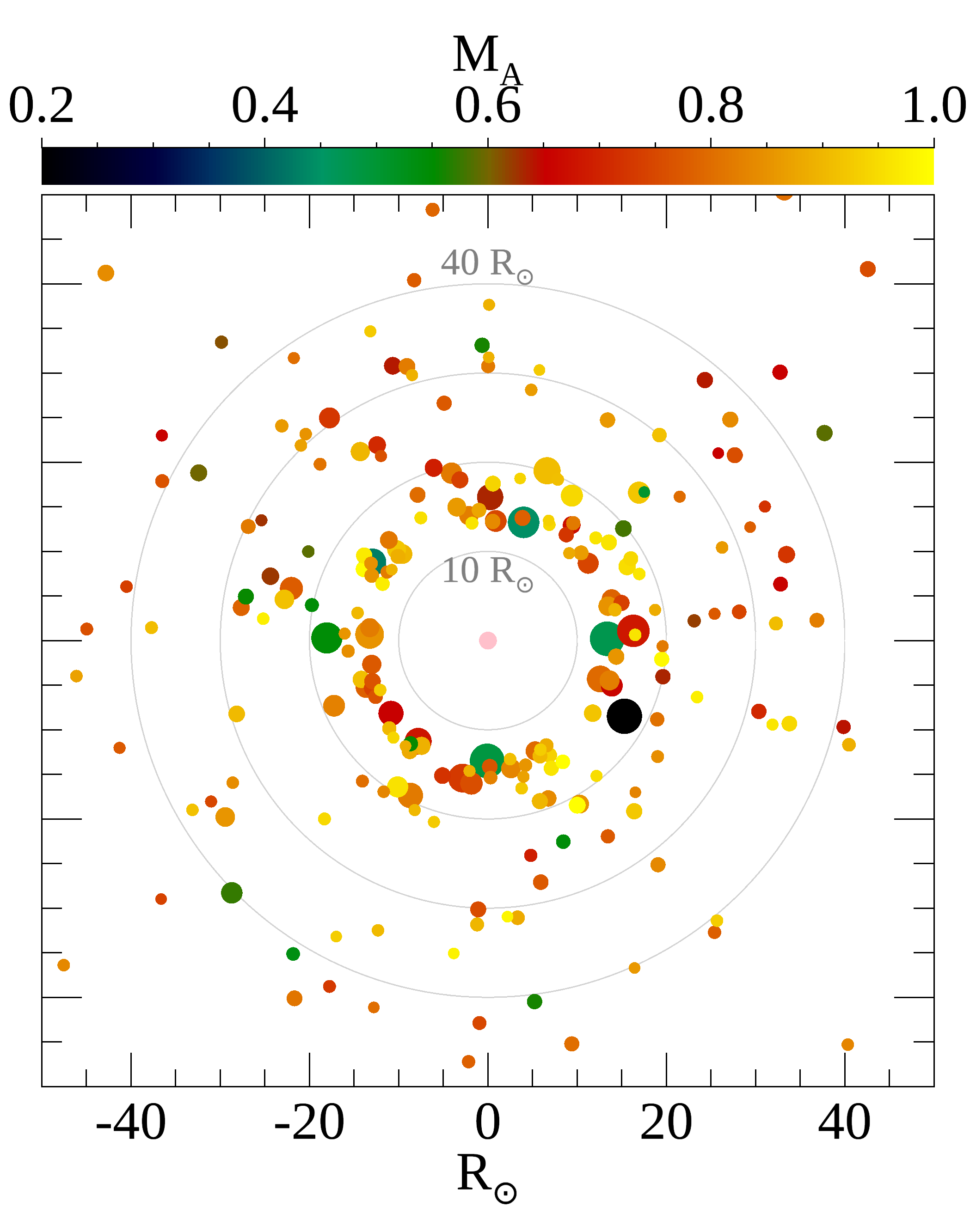}
    \includegraphics[width=.51\columnwidth]{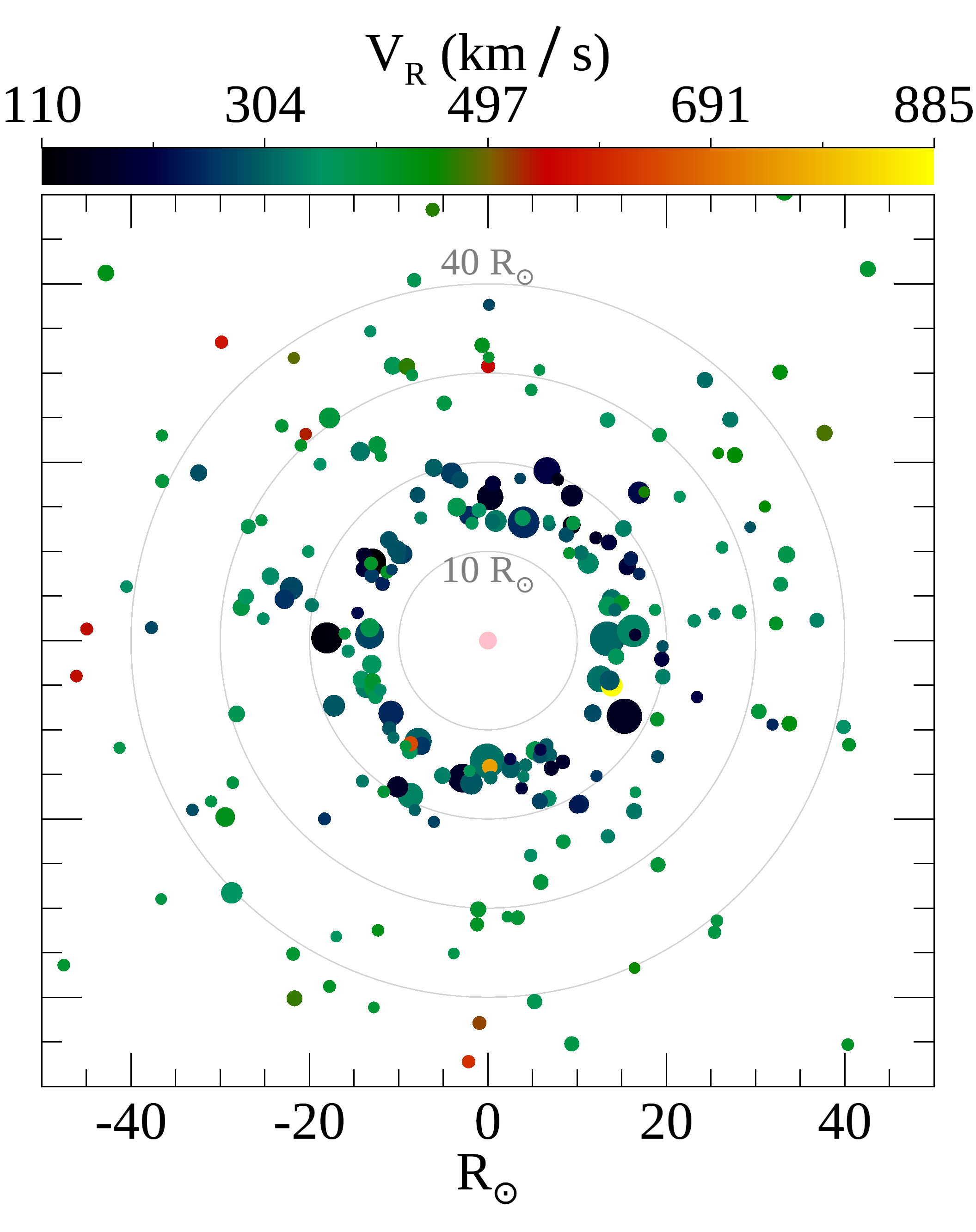}
    \includegraphics[width=.51\columnwidth]{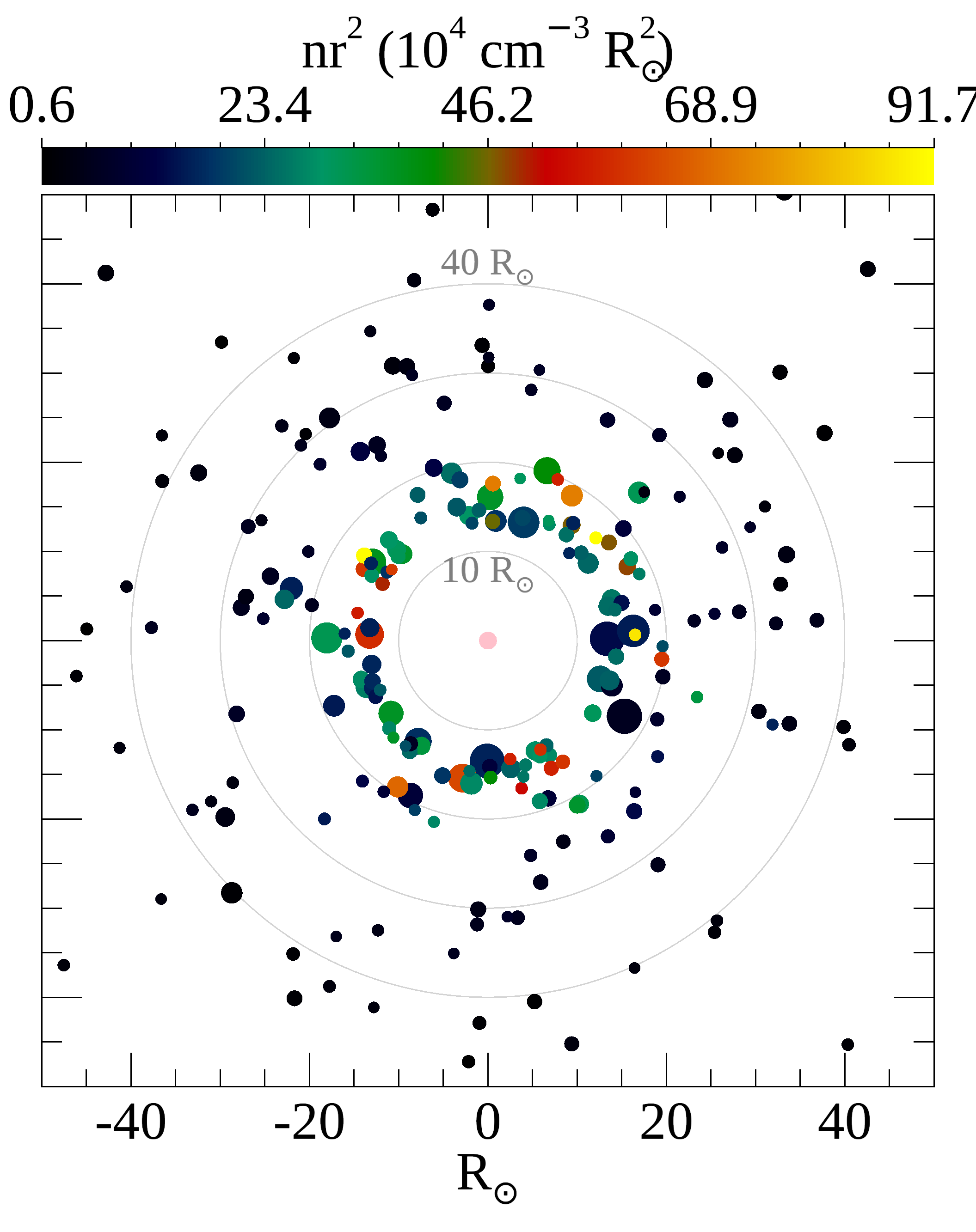}
    \includegraphics[width=.51\columnwidth]{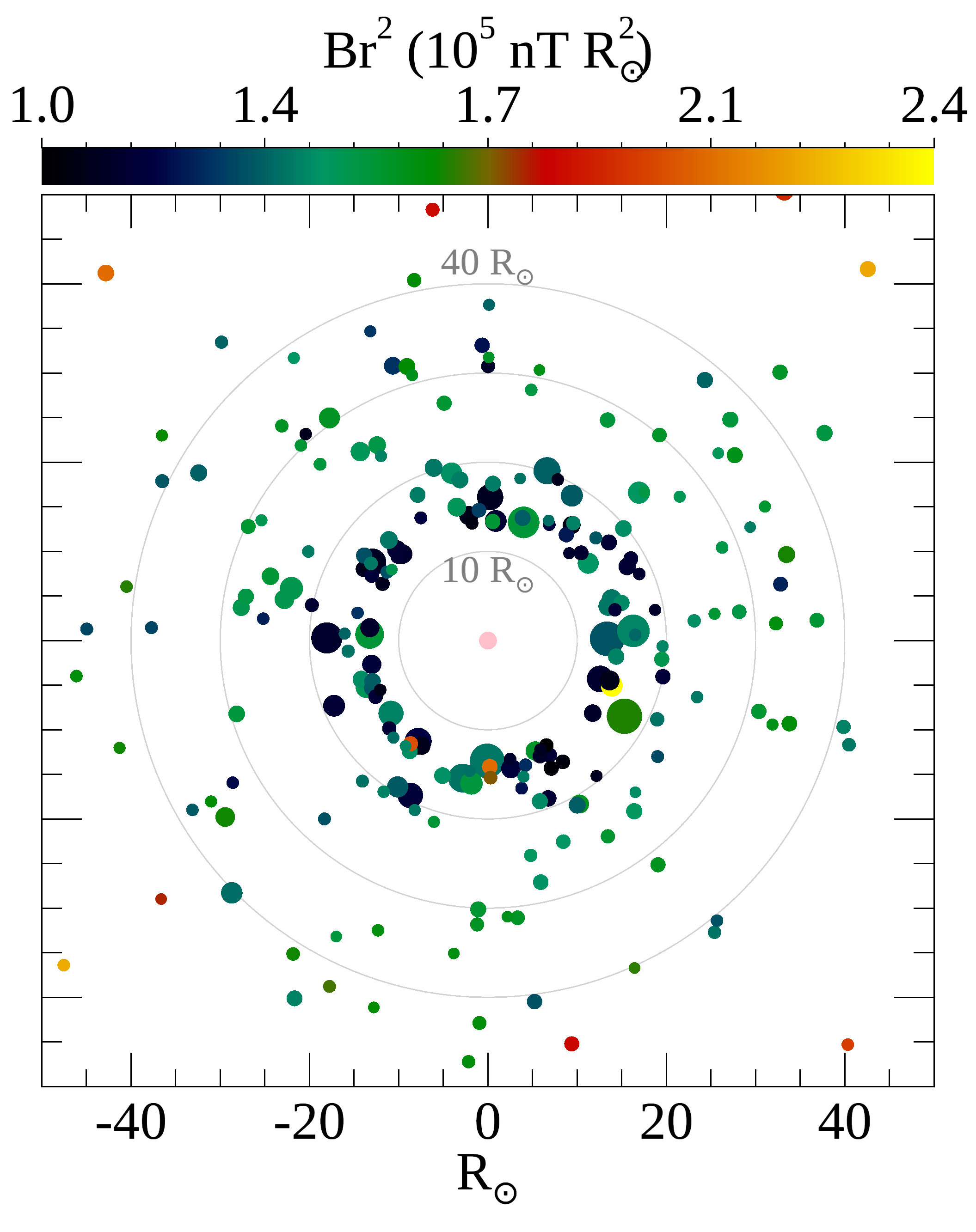}
    \caption{Bubble plots representing subAlfv\'enic intervals in PSP E8-E14, as described in Figure \ref{fig:bub1} caption. Diameter of circles represents natural log of duration of subA intervals, and colors represent different mean quantities for each interval (as labeled in respective colorbars): Alfv\'en Mach number \(M_A\), radial solar wind proton speed \(V_R\), \(nr^2\), and \(Br^2\), where \(n, B,\) and \(r\) are, respectively, proton density, magnetic field magnitude, and helioradius of respective intervals.}
    \label{fig:bub2}
\end{figure*}

In the final series of figures we present a visualization of the Alfv\'en zone as observed by PSP during E8-E14, via a series of so-called bubble plots (Figures \ref{fig:bub1} and \ref{fig:bub2}). For this purpose, we retain the helioradius of each interval but randomize their heliolongitudes, by assigning a random value for the latter drawn from a uniform distribution of longitudes between 0\degree~and 359\degree. Each circle represents a subA interval, where the center of the circle corresponds to the helioradius of the temporal midpoint of the interval. Diameters of circles represent the duration or spatial size of intervals, as specified, and are scaled proportionally to the smallest circle, which has arbitrary diameter and is \textit{not} to scale relative to the displayed heliocentric axes. This generates a distribution of subAlfv\'enic intervals/blobs in the ecliptic region (as defined by PSP's trajectory), and is therefore a (partial) visualization of the Alfv\'en zone in the observed young solar wind. \footnote{Randomization of longitudes enables simultaneous display of several PSP encounters. This representation complements the conventional style of plotting quantities such as \(M_A\) along PSP's trajectory during individual encounters \citep[e.g.,][]{badman2023JGR}. Plotting the PSP trajectory for several encounters in one figure would result in a cluttered and unclear visual.} 

The top panel of Figure \ref{fig:bub1} shows subA intervals with circle size and color representing interval duration. The speckled distribution above 30 \(\rs\) is evocative of the small subAlfv\'enic blobs in Figure \ref{fig:azone}. As in Figure \ref{fig:2dhist_1}, larger circles are seen below 20 \(\rs\), even as smaller circles endure, suggesting a highly patchy Alfv\'en zone. The middle panel of Figure \ref{fig:bub1} replaces interval duration with a rough estimate of spatial size: \(L_\text{subA} = \text{duration} \times V_\text{rel}\), where \(V_\text{rel}\) is the magnitude of the relative velocity between PSP and the solar wind. Here the solar wind and PSP velocities are averages over the interval.\footnote{We neglect \(T\) and \(N\) components of solar wind velocity since these are small relative to the \(R\) component, and obtaining reliable estimates of non-radial velocities from SWEAP data appears to be non trivial \citep{badman2023AGU_PSP_azimuthal}. Their neglect is not expected to modify our results in this case, namely, rough estimates of \(L_\text{subA}\).} \(L_\text{subA}\) ranges from 1-20 \(\times 10^6\) km, with a distribution similar to that of the duration. We note that the relative sizes of circles are not preserved exactly on transforming duration to spatial size. The bottom panel shows the natural logarithm of interval duration, reducing the extreme size differences in circles in the top and middle panels.

Figure \ref{fig:bub2} shows distributions of subAlfv\'enic intervals in a format similar to the last panel of Figure \ref{fig:bub1}, where circle diameter represents log (duration).\footnote{The logarithm of duration is chosen to represent circle size in order to enable visualization of the color of smaller blobs, and consequently to discern gradients in color.} The colors now represent four additional quantities of interest -- \(M_A\), and the quantities used to compute it, namely radial solar wind proton speed \(V_R\), proton density \(n\), and magnetic field magnitude \(B\). Each of these is computed as an average over an individual interval.\footnote{Here \(B = \langle(B_R^2 + B_T^2 + B_N^2)^{1/2}\rangle\), where \(\langle\cdot\rangle\) denotes an average over an interval.} The latter two quantities have been multiplied by the square of the helioradius of the respective interval, in order to reveal trends beyond their expected approximate \(r^{-2}\) dependence \citep[e.g.,][]{chhiber2021ApJ_psp}.

Examining the top left panel, any radial gradient in \(M_A\) is hard to discern (cf. Figure \ref{fig:2dhist_2}), but one can see that low \(M_A\) tends to be associated with large blob size. Low \(M_A\) can emerge from any combination of low \(V_R\), low \(n\), or large \(B\), and the three other panels examine these factors. The \(V_R\) panel shows that nearly all subA intervals have slow wind speeds, with only about 10 blobs having \(V_R > 500\) km/s. The speed doesn't appear to change much with \(r\), but larger blobs tend to have smaller speeds. Having corrected for the radial trend in \(n\) and \(B\), we observe that the ``innermost'' blobs \(r<20~\rs\) have a large variability in density, while the magnetic field is more uniform.

\section{Discussion and Conclusions}

In order to characterize the transition from sub- to super-Alfv\'enic flow in the young solar wind, we have used PSP data aggregated from its 8th to 14th solar encounters and identified around 220 subAlfv\'enic periods with duration \(\ge\) 10 minutes. The distribution of durations, heliocentric distances, and Alfv\'en Mach numbers of the intervals is consistent with the notion of a patchy, fragmented, and turbulent Alfv\'en zone, wherein the transition occurs over an extended range of helioradii. Comparison of observations with a 3D model Alfv\'en zone that includes turbulence effects \citep{Chhiber2022MNRAS} further supports this view. As PSP approaches the Sun, the frequency of both short and long duration intervals increases, with the latter associated with smaller \(M_A\). The model comparison suggests that, as of E14, PSP is yet to venture deep within the subAlfv\'enic domain of the corona; future observations, especially the planned perihelia below 10~\(\rs\), may observe long-duration subAlfv\'enic periods to the exclusion of short durations.

The patchy and fragmented nature of the Alfv\'en transition region may have implications for related phenomena. This includes the transfer of angular momentum from the Sun to the corona (and in stellar atmospheres in general), which is effective inside the Alfv\'en surface \citep{weber1967ApJ148} and has been shown to be influenced by turbulent stresses \citep{usmanov2018}. The propagation, reflection, and dissipation of Alfv\'en waves in this region \citep{verdini2009ApJ} could also be affected, resulting in complex wave characteristics, and
a stochastic variability of wave propagation \citep{Cranmer2023SoPh}. 

The list of  subAlfv\'enic intervals identified here 
will be employed in more detailed analyses of coronal plasma properties in forthcoming studies. The present results provide context for these, and also for NASA's planned PUNCH mission, which aims to perform global mapping of the Alfv\'en zone using remote imaging techniques \citep{DeForest2022IEEE,Cranmer2023SoPh}. In further development of the Alfv\'en zone model we plan to include a realization of explicit velocity fluctuations, in addition to the magnetic fluctuations considered in \cite{Chhiber2022MNRAS}; the correlation between the two will be constrained by the cross helicity distribution obtained from the turbulence transport model. This development may result in a more turbulent and fragmented Alfv\'en zone, which could further improve agreement with PSP observations.

\section*{Acknowledgements}
This research is partially supported by NASA under Heliospheric Supporting Research program grants 80NSSC18K1210, 80NSSC18K1648, and 80NSSC22K1639, Parker Solar Probe Guest Investigator program grant 80NSSC21K1765, LWS Science program grant 80NSSC22K1020, and the PUNCH project under subcontract NASA/SWRI N99054DS. This research has been supported in Thailand by the National Science and Technology Development Agency (NSTDA) and National Research Council of Thailand (NRCT): High-Potential Research Team Grant Program (N42A650868). Computing resources supporting this work were provided by the NASA High-End Computing (HEC) Program through the NASA Advanced Supercomputing Division at Ames Research Center. We acknowledge the PSP mission for use of the data, which are publicly available at \href{https://spdf.gsfc.nasa.gov/}{NASA Space Physics Data Facility}. PSP ephemeris was extracted from SPICE kernels obtained from \href{https://psp-gateway.jhuapl.edu/}{PSP Science Gateway}.

\section*{Data Availability}

Simulation data will be made available upon reasonable request to the corresponding author. PSP data are publicly available at the \href{https://spdf.gsfc.nasa.gov/}{NASA Space Physics Data Facility}.
 









\bsp	
\label{lastpage}
\end{document}